\documentclass[12pt]{article}
\usepackage{floatfig,epsfig}
\def\half{{\textstyle{1\over2}}}
\def\thalf{{\textstyle{3\over2}}}
\def\fhalf{{\textstyle{5\over2}}}

\def\lpmb#1{\mbox{\boldmath $#1$}}

\begin{document}
\initfloatingfigs
\begin{center}
{\Large {\bf Quark Models of Baryon Masses and Decays}}
  \\[\baselineskip]
{\large Simon Capstick}
  \\[\baselineskip]
{\large{\it Department of Physics, Florida State University\\
            Tallahassee, FL, USA 32306}}
  \\[2\baselineskip]
\end{center}
\vspace*{2.0cm}

\noindent
{\bf Abstract:}\\ {\small The description of baryon resonance masses
and decays in the constituent quark model is outlined, with emphasis
on the potential-model approach combined with the $^3P_0$
pair-creation model of strong decays. This approach allows the
estimation of branching fractions for baryon states missing in $\pi N$
elastic scattering analyses but expected to be present in
electromagnetic production. Prospects for the discovery of hybrid
baryons are discussed. A subjective list of the most important
corrections required to this approach is presented.  } \normalsize
\\[\baselineskip]

\section{Introduction}

Although many baryon resonances have been seen already in $\pi N\to
\pi N$, $\pi\pi N$, $\Lambda K$, {\it etc.}, many of these states
cannot be considered well known. Difficult multi-channel analyses are
required to find evidence for resonances from the data, which is
necessarily incomplete. Models also predict more states than have been
seen in the analyses, and there is a focused effort to discover
evidence for as many as possible of these `missing' states in new data
from the reactions $\gamma N$, $e^-N\to \pi N$, $\pi\pi N$, $\eta N$,
$\omega N$, $K\Lambda$, $K\Sigma$, {\it etc.}

The exploration of known and novel resonances which analysis of this
new data will allow has important consequences for our understanding
of the nature of low-energy hadronic states. The various QCD-based
models disagree on the spectrum and even the relevant degrees of
freedom and so the number of excitations. There is controversy about
the nature of the short-range interactions between the quarks, and the
presence of tensor and spin-orbit interactions. QCD also predicts gluonic
excitations which have the same quantum numbers as conventional
three-quark states, but which have not so far been seen. All of the
states are broad and often overlap. It is a unique challenge to
those building models of the fundamental excitations and of the
reactions to unravel this physics.

Such a program does not directly test QCD. Currently, {\it ab-initio}
approaches like lattice-QCD are finding good masses for ground-state
baryons within the quenched approximation, and
full-QCD calculations of these masses are
underway~\cite{lattice}. Calculations are also
underway of $P$-wave baryon masses. However, it will remain difficult
to predict with lattice QCD the masses and decay branches of the third
and fourth $P_{11}$ $N\half^+$ or $P_{13}$ $N\thalf^+$ excited states,
for example. The reason for the interest in these highly-excited
conventional states is that, as we shall see, establishing their
presence rules out models with strong diquark clustering, and the
lightest gluonic excitations are likely to have the same quantum
numbers.

What this program does test are QCD-based models, which are
based on interpreted consequences of QCD. These include potential
models, chiral perturbation theory, a collective model based on
spectrum-generating algebra, large-$N_c$ expansions, and recently
relativistic field-theoretic models based on QCD. This work focuses on
the description of baryon masses and decays using potential models.

\section{Potential models}

In this approach the effective degrees of freedom are constituent
quarks, treated symmetrically, unlike those in
quark-diquark~\cite{Goldstein:1988us} and collective~\cite{Bijker}
models. These constituent quarks move in two- and three-body
potentials. Initial formulations were non relativistic and based on
the Schr\"odinger equation~\cite{Isgur:1978xj}, although many authors
now use the relativistic kinetic energy~\cite{Glozman:1998xs} (with a
lower constituent quark mass) and some introduce relativistic
corrections~\cite{Capstick:1986bm} in the potentials. The dynamical
problem is solved by expanding the wave function in a large basis,
generally the harmonic-oscillator basis, which has many advantages. It
is also possible to use a basis which interpolates between the Coulomb
and linear eigenfunctions expected from the short and long-distance
behavior of the spin-independent potential,
respectively~\cite{Carlson:1983xi}. Once the Hamiltonian matrix has
been constructed, it can be diagonalized to find the spectrum and wave
functions, and the size of the basis increased until a reasonable
convergence of the spectrum is attained.

\begin{floatingfigure}{9.0cm}
\mbox{\epsfig{file=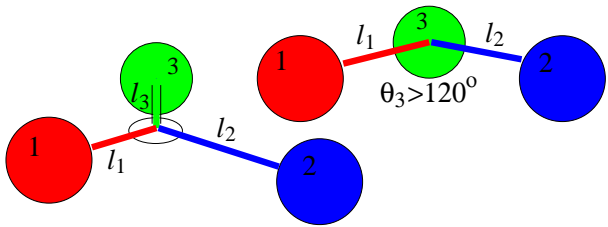,width=80mm}}
\caption{\small{String configurations leading to the confining
potential $V_{\rm conf}=b\sum_i l_i$ in baryons.}
}
\label{fig:string}
\end{floatingfigure}

A popular model of confinement in baryons uses flux-tubes or strings
attached to the quarks and to a junction, combined with the adiabatic
approximation, where the strings are assumed to adjust rapidly to the
quark motion. A potential is generated by minimizing the string length
for each set of quark positions. Two cases are illustrated in
Fig.~\ref{fig:string}, that where all of angles in the triangle made
by joining the quarks are less than 120$^\circ$, or where one of the
angles is larger than 120$^\circ$. In the first case the sum of the
string lengths is minimized when the angles at the junction are all
120$^\circ$, and in the second it is when the junction is over the
quark where the large angle resides. In both cases the confining
potential is the sum of the string lengths times the string tension
$b\simeq 1.0$ GeV/fm, and it is this potential which is used in modern
baryon spectrum calculations.

Quarks are known to exchange gluons with a coupling which is weak at
short distances.  The short distance potential between the quarks is
modeled using this one-gluon exchange interaction as a guide. An
alternate model (OPE) uses the exchange of an octet of pseudoscalar
mesons~\cite{Glozman:1996fu}, with the motivation that the
short-distance dynamics of constituent quarks should be governed by
dynamical chiral-symmetry breaking. The one-gluon-exchange interaction
is similar to the hyperfine interaction between the proton and the
electron in the hydrogen atom, with contact and tensor pieces,
although it is much stronger. It is responsible for the approximately
300 MeV splitting of the $\Delta(1232)$ and the nucleon. There are
also spin-orbit interactions implied by the one-gluon exchange
interaction, and spin-orbit interactions due to Thomas-precession in
the confining potential are present regardless of the source of the
short-distance interactions.

More recent models generally use the relativistic kinetic energy for
the quarks and a quark mass of about 220 MeV for the light quarks. The
relativized model~\cite{Capstick:1986bm} also applies
momentum-dependent relativistic corrections to the confining and
short-distance potentials. The wave functions are then expanded in a
large (often harmonic-oscillator) basis, and the resulting Hamiltonian
matrix is then diagonalized to find the spectrum and wave functions.

\begin{floatingfigure}{9.0cm}
\mbox{\epsfig{file=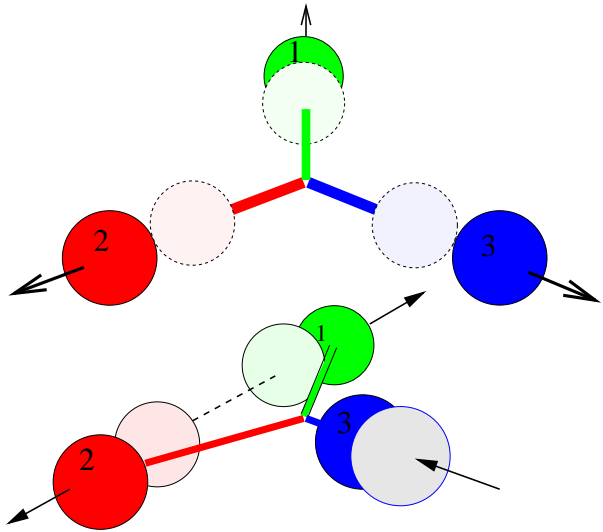,width=80mm}}
\caption{\small{Symmetric radially excited spatial state corresponding
to the Roper resonance, and one of the two mixed-symmetry radial
states.}
}
\label{fig:radial}
\end{floatingfigure}

As a relevant illustration of this technique, there are three
radially-excited harmonic oscillator spatial states, one in each of
the two relative coordinates and one unique to the three-body system
proportional to the dot product of the two relative coordinates
($\lpmb{\rho}\cdot \lpmb{\lambda}$). When combined into states of
definite exchange symmetry, these make up a symmetric state and a pair
of mixed-symmetry states, two of which are illustrated in
Fig.~\ref{fig:radial}. When these $L^P=0^+$ states are combined with
the mixed-symmetry quark-spin $S=\half$ and totally symmetric
$S=\thalf$ states, the result is five states nominally in the $N=2$
oscillator band (but note the potential is not harmonic). These are
two $N\half^+$ states, a $\Delta \half^+$ state, a $N\thalf^+$ state,
and a $\Delta \thalf^+$ state. Some of these states have been seen on
pion-nucleon elastic scattering. These are the Roper resonance
$N\half^+(1440)$, its analog $\Delta \thalf^+(1600)$, and
$N\half^+(1710)$.

Decay models based on the elementary-meson emission~\cite{KI} or the
$^3P_0$ pair-creation model~\cite{Capstick:1993th} can explain why the
other states have not been seen--they have small $N\pi$ couplings--and
can predict alternate final-state channels in which they are likely to
be seen. In one-gluon exchange models the Roper resonance is lighter
than the other positive-parity excited states due to a strongly
negative contact interaction like that in the proton, and it is also
split from the other positive-parity states by the anharmonic nature
of the confining potential~\cite{Isgur:1978xj}. When treated properly,
without resort to perturbation theory in the large anharmonic
perturbation, the Roper resonance becomes much lighter than the other
positive parity excited states, but not light enough to explain its
position in the spectrum below the lowest (at 1520 MeV) of the
negative-parity non-strange excited states. The next $P_{11}$ state,
$N\half^+(1710)$, is in roughly the right position in the spectrum.

Models with OPE-based short-distance interactions explain the low
position of this state as due to the flavor-dependence of the
resulting contact interaction~\cite{Glozman:1996fu}.  However, the
Roper resonance is very broad (roughly 350 MeV width), and neither of
these approaches takes into account shifts in its mass due to
self-energy corrections from baryon-meson loops, which can naively be
expected to be of the order of the width. It is therefore not clear
whether 100 MeV discrepancies in any spectrum will allow conclusions
to be made until this rather difficult problem is consistently dealt
with, for all the states in the spectrum.

\begin{figure}[bt]
\begin{center}
\mbox{\epsfig{file=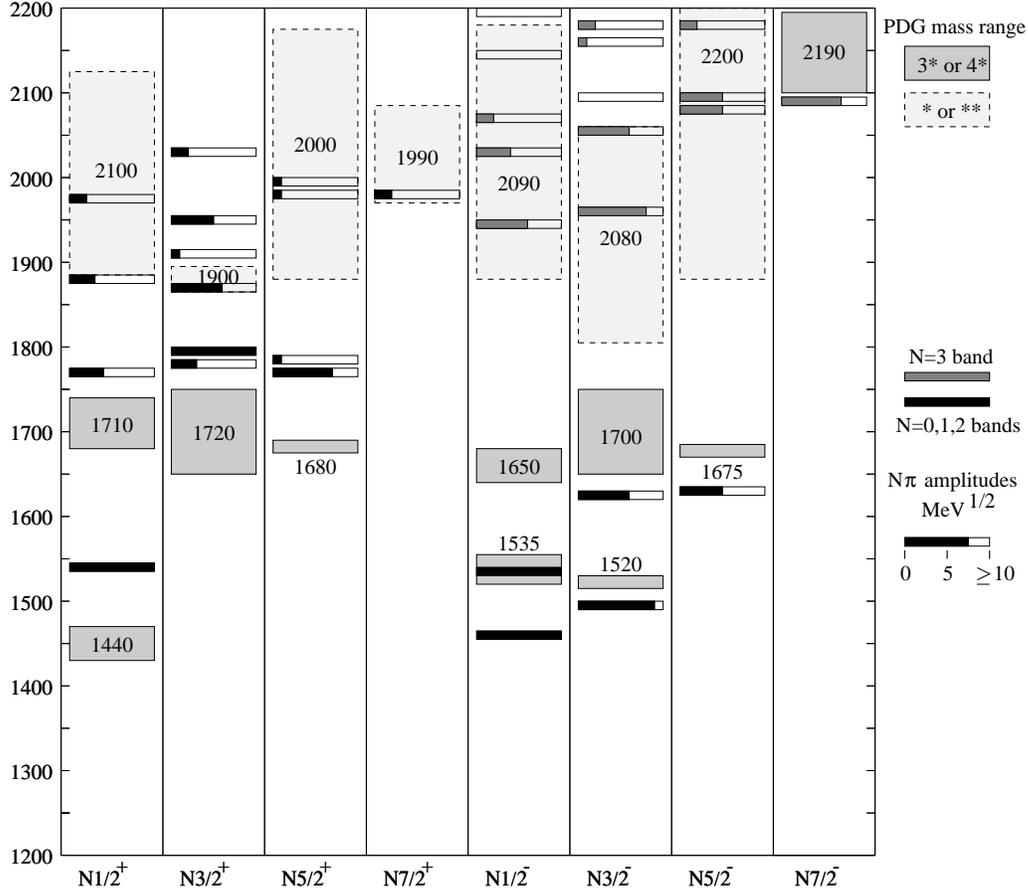,width=140mm}}
\caption{\small{Calculated masses and $N\pi$ decay amplitudes for
nucleon resonances below 2200 MeV from
Refs.~\protect\cite{Capstick:1986bm,Capstick:1993th}, compared to the
range of central values for resonances masses from the
PDG~\protect\cite{Groom:2000in}, which are shown as boxes. The boxes
are lightly shaded for one and two star states and heavily shaded for
three and four star states. Predicted masses are shown as a thin bar, with
the length of the shaded region indicating the size of the
$N\pi$ amplitude.}}
\label{fig:Nexpthr}
\end{center}
\end{figure}

\begin{figure}[bt]
\begin{center}
\mbox{\epsfig{file=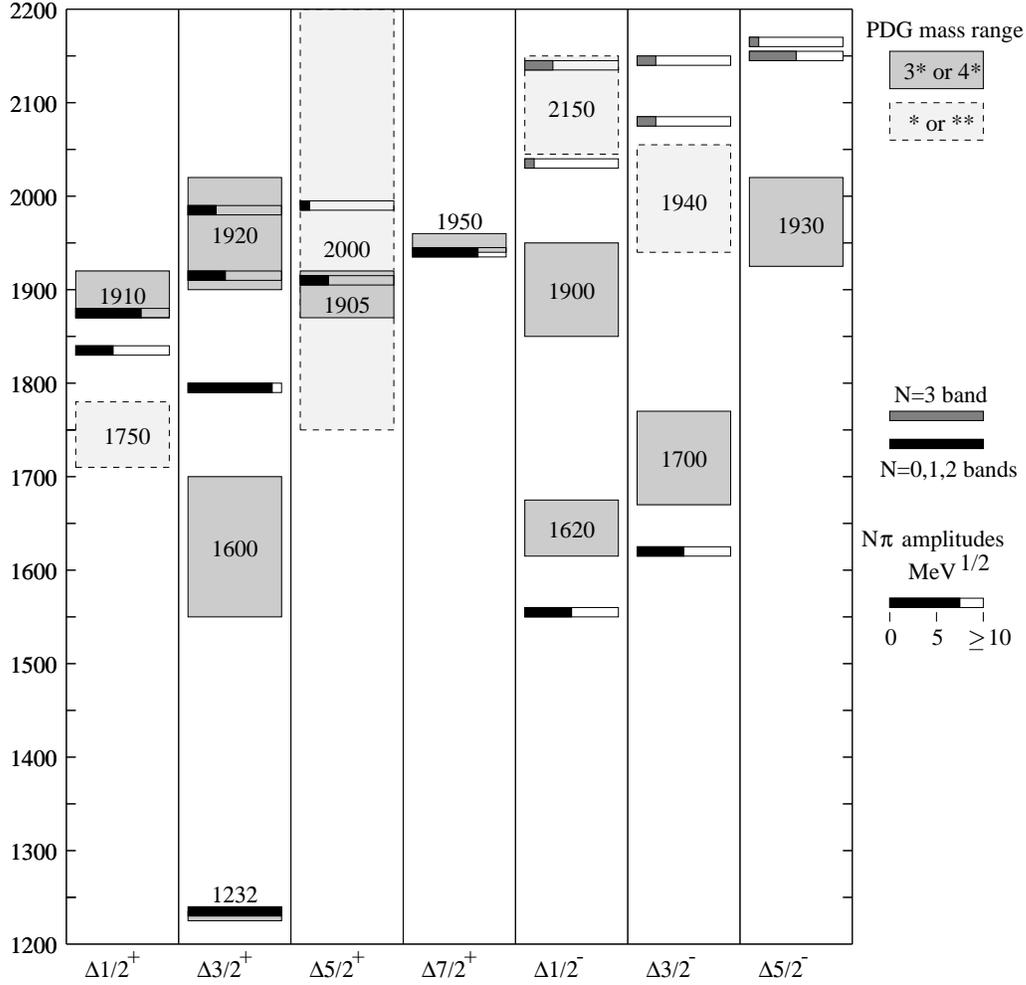,width=140mm}}
\caption{\small{Model masses and $N\pi$ decay amplitudes for $\Delta$
resonances below 2200 MeV from
Refs.~\protect\cite{Capstick:1986bm,Capstick:1993th}, compared to
the range of central values for resonances masses from the
PDG~\protect\cite{Groom:2000in}. Caption as in
Fig.~\protect\ref{fig:Nexpthr}.}}
\label{fig:Dexpthr}
\end{center}
\end{figure}

Masses for nonstrange baryon states below 2200 MeV calculated in this
way are shown in Figures~\ref{fig:Nexpthr} and~\ref{fig:Dexpthr},
along with $N\pi$ decay amplitudes~\cite{Capstick:1993th} (their
squares give the $N\pi$ partial widths) for each state. The calculated
masses are shown as thin bars, and the length of the shaded part of
each bar is proportional to the $N\pi$ decay amplitude strength. Also
shown in these figures are boxes showing the range in the central
value of the mass of resonances quoted by the Particle Data
Group~\cite{Groom:2000in} (along with their best estimate of the
mass), which are compiled from partial-wave analyses of mainly $N\pi$
elastic and inelastic scattering data. It is clear that low-lying well
separated states with substantial $N\pi$ widths are those likely to
have been seen in the analyses, and that model states with small
$N\pi$ couplings are either poorly established resonances, or not
present at all in the analyses. Models which describe the degrees of
freedom in baryons at low energy as quarks and diquarks have fewer
degrees of freedom and so fewer excitations, so the discovery of
additional positive parity excited states (and the confirmation of
some existing states) predicted by models which treat the quarks
symmetrically could rule out this possibility.

Further insight into the nature of the short-distance interactions
between quarks can be found from a consideration of the nature of the
spin-orbit and tensor interactions in baryons. Although the
cancellation of the two-body parts of the one-gluon exchange (OGE)
spin-orbit interactions and those arising from Thomas precession in
the confining potential can be arranged~\cite{Isgur:1978xj},
spin-orbit splittings are still too large in non relativistic
OGE-based models. In the relativized model it is shown that it is
possible to have small spin-orbit interactions in baryons due to
relativistic effects~\cite{Capstick:1986bm}. This problem is not
solved by the elimination of the spin-orbit forces due to the
short-distance interactions, such as in OPE-based models, as this to a
large extent is required to cancel against those from the confining
potential~\cite{Isgur:1999jv}.

Pseudoscalar exchange (OPE) models of the short-distance interactions
between quarks have tensor interactions which are very different from
those arising from OGE. These interactions have important consequences
for the tensor mixings between states such as the $N\half^-(1535)$ and
$N\half^-(1650)$, which are strong. The mixings arising from OGE
explain why the lower of these two states has a large
branch~\cite{Isgur:1978xj} to $N\eta$, which is not the case for
mixings found using OPE. Some form of vector exchange, such as gluon
exchange or the exchange of two pions in the form of a vector
meson~\cite{Wagenbrunn:2000sg}, is required to describe the details of
the strong decays of these states.

Finally, it is clear that OPE-based descriptions of the baryon
spectrum cannot reproduce the reasonable unification of baryon and
meson physics provided by OGE-based models. They cannot produce
hyperfine mixing in mesons where it is needed, and imply unphysical
OZI-violating mixings in isoscalar mesons~\cite{Isgur:1978xj}. It is,
perhaps, natural~\cite{Glozman:2000vd} to exclude a description of
meson states in a model where bound quarks interact by exchanging
those same meson states.

Interesting and different physics can be accessed by studying states
containing one ($\Sigma$ and $\Lambda$) or more ($\Xi$) strange
quarks. The presence of the strange quark breaks degeneracies present
in the nonstrange spectrum and so there are more negative-parity
($L=1$) excited states. Our knowledge of the spectrum from
partial-wave analyses is limited to the ground states $\Lambda$,
$\Sigma$, and $\Sigma^*(1385)$, and about half of the negative-parity
excited states and a few of the low-lying positive-parity states
predicted by constituent quark models. The spin-orbit partners
$\Lambda(1405)\half^-$ and $\Lambda(1520)\thalf^-$ provide an
interesting challenge for quark models, which need to suppress
spin-orbit splittings elsewhere in the spectrum to agree with the
analyses. These states are predicted to be roughly degenerate in
potential models. Because of this poor description in the constituent
quark model they been described as $\bar{K}N$ bound states. It is also
possible that their large mass splitting arises from corrections to
their masses from $qqq(\bar{q}q)$ configurations, as the lightest
state lies just below threshold for decay to $\bar{K}N$. Note that
OPE models can be adjusted to accommodate this splitting with
explicitly flavor-dependent contact interactions (not just through the
mass dependence of the hyperfine interaction, as in OGE-based models).

There is information on only a few excited $\Xi$ states extracted from
the data, which include the ground states $\Xi$ and $\Xi^*$, and a few
negative-parity excited states. For this reason it is important to
search for information on excited $\Xi$ states. There is recent
evidence from Jefferson Laboratory of a signal for $\Xi$ states in
$e^-p\to \Xi KK$.

\section{Hybrid baryons}

The discovery of hybrid baryons is complicated by the fact that the
quantum numbers of baryons with excited glue are the same as those of
conventional three-quark excitations. Using the flux-tube picture of
the confining potential described above, it is
possible~\cite{Capstick:1999qq} to describe light hybrid baryons as
states where the quarks move in a potential described by an excited
state of the glue. This is very different from earlier $qqqg$
models~\cite{Barnes:1983fj} of hybrid baryons, where a gluon and three
quarks are confined to a bag.

\begin{floatingfigure}{8.0cm}
\mbox{\epsfig{file=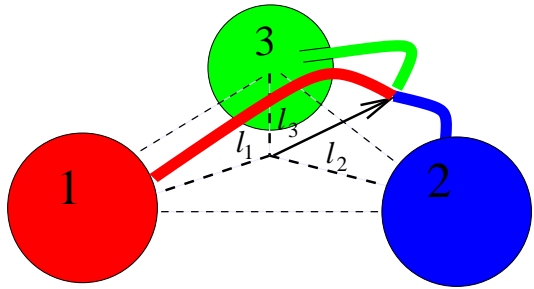,width=70mm}}
\caption{\small{General motion of the flux tubes (strings) in a hybrid
baryon.}}
\label{fig:hybrid}
\end{floatingfigure}

For fixed positions ${\bf r}_i$ of the three quarks, the flux tubes
are allowed to move. This is modeled by the motion of the junction,
and by transverse motion of the strings relative to their equilibrium
directions, as in Fig.~\ref{fig:hybrid}. By means of an analytic
calculation with a bead at the junction and a single bead on each of
the three strings, it was shown that a reasonable approximation to the
ground-state and first-excited-state energy of the string for each
quark position can be found by examining only the junction motion. The
strings simply follow the motion of the junction. The adiabatic
potentials corresponding to the ground state of the string, $V_B({\bf
r}_1, {\bf r}_2, {\bf r}_3)$, and the the first excited state of the
string, $V_H({\bf r}_1, {\bf r}_2, {\bf r}_3)$, are found by a
four-parameter variational calculation, for all quark positions. 

The difference of these potentials is then added to the Hamiltonian
for the three quark problem, and the energies of the lightest hybrid
baryons are then found by solving the three-quark dynamical problem as
before, but with the excited string Hamiltonian. It can be shown that
the string excitations have orbital angular momentum and parity $1^+$,
so that when combined with the lowest energy quark motion in the new
potential, which is still $L=0$ and $S=\half$ or $\thalf$, the
resulting lowest-lying states are $N\half^+$ and $N\thalf^+$ at
roughly 1870 MeV, with a model error of about $\pm 100$ MeV. There are
also $\Delta\half^+$, $\Delta\thalf^+$, and $\Delta\fhalf^+$ states at
about 2075 MeV with the same estimated error. In the bag
model~\cite{Barnes:1983fj} there is a $N\fhalf^+$ excited state and no
$\Delta\fhalf^+$ state, and the lightest states are at about 1500 MeV,
significantly lighter than these flux-tube model calculated masses.

Referring back to Figs.~\ref{fig:Nexpthr} and~\ref{fig:Dexpthr}, we see
that these new states are predicted by this flux-tube model to lie in
a region of the spectrum where there are several missing conventional
three-quark excitations. Given that these states are likely to mix
with the conventional excitations (through corrections to the adiabatic
approximation outlined above), it is important that a careful and
detailed analysis of these partial waves is carried out with an eye
towards establishing more states than are possible without exciting
the flux-tubes. It may also be possible to distinguish hybrids based
on their electromagnetic excitation amplitudes~\cite{Li:1992yb} and
strong-decay signatures.

\section{Baryon strong decays}

It is not enough to find a state in a model with a mass similar to one
seen in analyses of scattering experiments (and so listed by the
Particle Data Group). In addition, it is necessary to explain why a
state should be seen, usually (for nonstrange states) in $N\pi\to X$,
and conversely why states present in the model which have not been
seen are missing in the analyses. Popular models of these strong
couplings are based on string breaking by the creation of a $q\bar{q}$
pair with vacuum quantum numbers. This model is referred to as the
$^3P_0$ model~\cite{LeYaouanc:1978bf}, as a such a quark and antiquark
necessarily have relative $L=1$ and $S=1$ combined to
$J^{PC}=0^{++}$. Models of meson decays preferentially create this
pair along the string connecting the quarks.

Progress has been made in understanding this model at the microscopic
level~\cite{barnes}. Certain meson decays can result in two relative
angular momenta for the final state mesons ($S$ and $D$-waves, for
example), and the ratio of the amplitudes for these decays depends
minimally on anything but the structure of the decay vertex. The model
of Ackleh, Barnes and Swanson (ABS) creates the quark pair by an interaction
with a quark in the decaying meson of the same kind as that which is
responsible for that meson's structure. ABS show that
this interaction resembles $^3P_0$ and that the comparison with these
ratios favors $^3P_0$ over the creation of a pair with gluon ($^3S_1$)
quantum numbers, for example.

The application of the $^3P_0$ model to baryon strong decays explains
why some of the states predicted by OGE-based models are missing in
analyses of the scattering data, which is because they have small
couplings to the $N\pi$ channel. Such selection rules may not apply to
electromagnetic production of such states, and this model can be used
to predict the strongest decay channels for these missing states. This
analysis depends not only on the form of the decay operator but also
on the model wave functions used for the baryons. A two-parameter model
fit to $N\pi$ decays has been used~\cite{CR2} to predict the $\Delta\pi$,
$N\rho$, $N\eta$, $N\eta^\prime$, $N\omega$, $N(1440)\pi$,
$\Delta(1600)\pi$, $\Delta\eta$, and $\Delta\omega$ decays of excited
baryon states. In addition, decays to $YK$ states have been examined,
where $Y$ is a hyperon $\Lambda$, $\Sigma$, $\Sigma^*(1385)$,
$\Lambda(1405)$, or $\Lambda(1520)$, and $K$ is a kaon or $K^*$. These
calculations show that it should be possible to discover `missing'
baryon states in several of these channels, especially those with
thresholds in the region where these excited positive-parity states
lie.

\begin{table}
\begin{tabular}{|r|r|r|r|r|r|r|r|r|r|r|r|r|r|} 
\hline
State & $\Gamma$ &  $N\pi$ &  $\Delta\pi$ &  $N\rho$ &  $N^*\pi$ & 
  $\Delta^*\pi$ &  $N\eta$ &  $N\omega$ & 
  $\Delta\eta$ & $\Lambda K$ & $\Sigma K$ & $\Sigma^*K$ \\ 
\hline
 $[N     \frac{1}{2} ^+]_4 (1880)$ &  150 &     .05 &     .49 &
.03 &     .00 &     .00 &     .18 &     .14 &     .00 & .00 & .09 & .00\\
 $[N     \frac{1}{2} ^+]_5 (1975)$ &   50 &     .08 &     .47 &
.14 &     .01 &     .00 &     .00 &     .22 &     .00 & .03 & .01 & .04\\
 $[N     \frac{3}{2} ^+]_2 (1870)$ &  190 &     .20 &     .12 &
.02 &     .01 &     .02 &     .26 &     .11 &     .00 & .00 & .26 & .00\\
 $[N     \frac{3}{2} ^+]_3 (1910)$ &  390 &     .00 &     .75 &
.03 &     .01 &     .01 &     .00 &     .17 &     .00 & .00 & .02 & .01\\
 $[N     \frac{3}{2} ^+]_4 (1950)$ &  140 &     .12 &     .43 &
.11 &     .00 &     .01 &     .00 &     .28 &     .00 & .03 & .01 & .01\\
 $[N     \frac{3}{2} ^+]_5 (2030)$ &   90 &     .04 &     .57 &
.15 &     .00 &     .01 &     .00 &     .16 &     .00 & .01 & .00 & .06\\
 $[N     \frac{5}{2} ^+]_2 (1980)$ &  270 &     .01 &     .89 &
.02 &     .00 &     .01 &     .00 &     .03 &     .00 & .00 & .00 & .05\\
 $[N     \frac{5}{2} ^+]_3 (1995)$ &  190 &     .00 &     .51 &
.33 &     .00 &     .01 &     .04 &     .08 &     .00 & .00 & .00 & .02\\
 $[N     \frac{7}{2} ^+]_1 (2000)$ &   50 &     .13 &     .53 &
.03 &     .00 &     .02 &     .21 &     .06 &     .00 & .00 & .02 & .00\\
 $[\Delta\frac{1}{2} ^+]_1 (1835)$ &  310 &     .05 &     .63 &
.20 &     .01 &     .01 &     .00 &     .00 &     .07 & .00 & .03 & .00\\
 $[\Delta\frac{3}{2} ^+]_4 (1985)$ &  220 &     .05 &     .44 &
.25 &     .00 &     .00 &     .00 &     .00 &     .17 & .00 & .05 & .03\\
 $[\Delta\frac{5}{2} ^+]_2 (1990)$ &  350 &     .00 &     .56 &
.10 &     .00 &     .00 &     .00 &     .00 &     .28 & .00 & .00 & .05\\
\hline
\end{tabular}
\caption{\small Total widths $\Gamma$ and branching fractions for
missing nonstrange baryons, from
Refs.~\protect\cite{Capstick:1993th,CR2}. States are labeled by their
spin, parity, principal quantum number, and model
mass~\protect\cite{Capstick:1986bm}. Here $N^*\pi$ means
$N\half^+(1440)\pi$, $\Delta^*\pi$ means $\Delta\thalf^+(1600)\pi$,
and $\Sigma^*K$ means $\Sigma\thalf^+(1385)K$. Branching fractions are
upper bounds, as discussed in the text.}
\label{table:missing}
\end{table}

The total widths and branching fractions to nonstrange and strange
final states calculated in this way for missing and poorly established
nonstrange baryons in the low-lying (nominally $N=2$)
positive-parity bands are shown in
Table~\ref{table:missing}. Branching fractions to $\Delta\omega$,
$N\eta^\prime$, $\Lambda\half^-(1405)K$, and $\Lambda(1520)\thalf^-K$
are omitted as they are never larger than 0.01. Note that these
branching fractions should be considered as upper bounds, as there are
final states which may have substantial branches, such as $N^*\pi$,
where $N^*$ is a negative-parity excited state, which are not
included in the total width. Model results for branching fractions are
likely to be more reliable than those for the total widths, as they
are ratios and some model uncertainties cancel
out. Table~\ref{table:missing} shows clearly that these states have
small $N\pi$ branches and that they may be easy to see in $N\pi\pi$
final states such as $\Delta\pi$ and $N\rho$, as in other channels
like $N\eta$~\cite{Svarc} and $N\omega$. Branching fractions into
strange final states are predicted to be somewhat smaller, with a few
exceptions, but information on these channels can be considered
complementary to that of non-strange final states, and so will likely
be very useful.

\section{Conclusions}

One-gluon-exchange based potential models, when combined with the
$^3P_0$ strong decay model, adequately describe diverse features of
excited baryon and meson physics. They have been applied to the
spectra, strong decays, and electromagnetic couplings of baryons such
as photocouplings and electroproduction amplitudes (not mentioned
here). The results of these calculations show reasonable agreement
with resonance properties extracted from analyses of data.

These models are not covariant, although they include relativistic
corrections, and electroproduction amplitudes have been calculated with
light-cone based relativistic models. We need new relativistic
structure models based on QCD. More importantly, constituent quark
models generally ignore Fock-space components of baryons beyond
simple $qqq$ configurations. This is equivalent to the quenched
approximation in lattice QCD, or to assuming that baryon-meson
intermediate states have no effects on the physics of the baryon
resonances. This is obviously an important missing ingredient, given
that none of these states are narrow. In addition, these models need
to take into account excitations of the glue.

The parallel development of a detailed and comprehensive data analysis
program is required to extract meaningful results for new and poorly
established baryons from exciting new data coming from several
sources. These results will certainly challenge our understanding of
the physics of excited baryons, and may lead to a new level of
understanding of the low-energy features of QCD.

\end{document}